\documentclass[12pt]{article}
\usepackage[english]{babel}
\usepackage[utf8x]{inputenc}
\usepackage{amsmath}
\usepackage{graphicx}
\usepackage{etoolbox}
\usepackage{ragged2e}
\usepackage{changepage}
\usepackage{xcolor}
\usepackage{color,soul}
\usepackage{hyperref}
\usepackage{titlesec}
\usepackage[parfill]{parskip}
\usepackage[margin=1in]{geometry}
\usepackage{times}
\usepackage{float}
\usepackage[numbers,super]{natbib}
\usepackage{censor}
\usepackage{subcaption}

\titleformat*{\section}{\fontsize{13}{12}\selectfont\bfseries}
\titleformat*{\subsection}{\fontsize{12}{12}\selectfont\bfseries}
\titleformat*{\subsubsection}{\fontsize{11}{12}\selectfont\bfseries}

\title{\large \textbf{Integrating A.I. in Higher Education: Protocol for a Pilot Study with 'SAMCares: An Adaptive Learning Hub'}} 

\author{
Syed Hasib Akhter Faruqui$^{1, *}$, Nazia Tasnim$^{2}$, Iftekhar Ibne Basith$^{1}$,\\ Suleiman Obeidat$^{1}$, Faruk Yildiz$^{1}$\\ \\
$^{1}$ Engineering Technology, Sam Houston State University \\
$^{2}$ College of Education, University of Texas at Austin \\
$^{*}$ Corresponding Author
}

\date{} 

\makeatletter 
\patchcmd{\@maketitle}{\begin{center}}{\begin{adjustwidth}{0.5in}{0.5in}\begin{center}}{}{}
\patchcmd{\@maketitle}{\end{center}}{\end{center}\end{adjustwidth}}{}{}
\makeatother

\begin{document}
\justifying
\maketitle
\thispagestyle{empty}

\section*{Abstract}
Learning never ends, and there is no age limit to grow yourself. However, the educational landscape may face challenges in effectively catering to students' inclusion and diverse learning needs. These students should have access to state-of-the-art methods for lecture delivery, online resources, and technology needs. However, with all the diverse learning sources, it becomes harder for students to comprehend a large amount of knowledge in a short period of time. Traditional assistive technologies and learning aids often lack the dynamic adaptability required for individualized education plans. Large Language Models (LLM) have been used in language translation, text summarization, and content generation applications. With rapid growth in AI over the past years, AI-powered chatbots and virtual assistants have been developed. This research aims to bridge this gap by introducing an innovative study buddy we will be calling the ‘\textit{SAMCares}’. The system leverages a Large Language Model (LLM) (in our case, LLaMa-2 70B as the base model) and Retriever-Augmented Generation (RAG) to offer real-time, context-aware, and adaptive educational support. The context of the model will be limited to the knowledge base of Sam Houston State University (SHSU) course notes. The LLM component enables a chat-like environment to interact with it to meet the unique learning requirements of each student. For this, we will build a custom web-based GUI. At the same time, RAG enhances real-time information retrieval and text generation, in turn providing more accurate and context-specific assistance. An option to upload additional study materials in the web GUI is added in case additional knowledge support is required. The system's efficacy will be evaluated through controlled trials and iterative feedback mechanisms. By interlinking Artificial Intelligence and assistive technology in an educational setting, this project aspires to advance personalized learning experiences for students, making meaningful strides in inclusive education.

\noindent\textbf{Keywords:} SAMCares, Large Language Model, Adaptive Learning, Interactive Learning, Retriever Augmented Generation

\section{Introduction}
\label{Section:Introduction}
The recent advancement in science and technology has transformed the field of higher education \cite{altbach2019trends}, bringing a paradigm shift in both teaching methodologies and learning experience. This trend can be observed globally from both students' and educators' perspectives. With the development of innovative educational platforms like adaptive learning platforms, virtual and augmented reality (VR/AR) training, and online courses, students get to experience an immersive and interactive learning experience \cite{de2010learning}. While traditional classroom settings primarily rely on direct instruction and passive learning, these new educational technologies foster an active learning environment where students are engaged and empowered to take charge of their education \cite{ruiz2023empowering}. 

\noindent However, these tools often feature pre-set training materials with certain limitations. These learning platforms, while advanced, generally come with pre-determined answers (online courses) and environments (VR/AR) that have been meticulously developed beforehand by a team of experts/educators \cite{chen2020artificial}. This structure can sometimes be restrictive, especially when learners pose creative or unconventional questions that deviate from the pre-set programming of the system or the scope of the lectures. The rigid framework of these tools means they are not always interpretive or flexible enough to accommodate queries that fall outside their programmed knowledge base or instructional design. Thus, this has led to a growing interest in more innovative and adaptive solutions in education technology that is both adaptive and can enhance the learning experience and outcomes for students in the field \cite{aggarwal2023integration}. The solutions should be capable of explaining concepts/topics to students in a manner that is responsive to their (students/learners) unique learning styles and adaptable to the diverse and sometimes unpredictable range of questions that a student might pose. 
These questions depend on the learners themselves. For example, a learner may have varying learning styles, such as visual, auditory, or reading/writing preferences, which can influence how they absorb and process information. Some may prefer hands-on example focused activities, while others may prefer reading text books to comprehend complex concepts. Additionally, cognitive abilities like memory, attention span, and problem-solving skills can influence how learners engage with educational materials. For example, some students may struggle with working memory, requiring additional support to retain information, while others may excel in logical reasoning, quickly grasping abstract concepts. Some who are anxious or lack confidence may require additional practice examples to build their self-efficacy, while those who are highly motivated may benefit from more challenging tasks to maintain their engagement. Furthermore, linguistic diversity must also be acknowledged, considering language preferences. Non-native English speakers may require additional language support to comprehend complex texts.
The ideal technology would be able to comprehend these conditions, interpret the knowledge and provide personalized and context-aware explanations similar to a human instructor. This level of adaptability would significantly enhance the learning experience, making it more engaging, effective, and tailored to individual students' needs.

\noindent In recent years, advances in artificial intelligence (AI), machine learning, and deep learning have revolutionized not only our day-to-day life but also industries ranging from businesses to finances to manufacturing facilities \cite{javaid2022artificial}. Among advances in AI, natural language processing stands out as a particularly transformative technology, especially in the context of higher education. Generative AI (GenAI) is an extension of NLP, which has shown how the most potential \cite{irugalbandara2023scaling}. GenAI models like Generative Pre-Trained Transformer (GPT) \cite{floridi2020gpt, zhang2023survey}, LLaMa \cite{touvron2023llama}, Mistral \cite{jiang2023mistral} etc. are some of the famous predictive generation language models that are currently being used to simulate human conversation, either written or spoken. A large amount of publicly collected data was used to train these models to process and produce text that is human-like. Interestingly, when prompted properly, these models have the ability to generate convincing, creative, and intelligent texts on a wide array of topics \cite{kung2023performance, kasneci2023chatgpt, sajja2023artificial}.
This capability of GenAI models opens up new possibilities in educational settings. Their application can range from assisting in creating educational content and personalized learning materials to offering sophisticated tutoring systems that can interact with students in a natural, conversational manner. These models can be particularly beneficial in enhancing engagement in e-learning platforms, where personalized, adaptive, context-dependent, and interactive content are key to student success. Thus offering a unique opportunity to reimagine the traditional educational model. By leveraging the abilities of GenAI models, educators and developers can create dynamic learning environments more aligned with individual learners' needs and preferences \cite{sajja2023artificial, grassini2023shaping}. In literature, several works have tried integrating these tools into their educational ecosystem. 
To offer interactive educational support to students in both text and voice mode, Lee et al. investigated the use of GenAI tools \cite{lee2022impacts}. Tai and Chen utilized a combination of decision tree hierarchies and natural language processing to provide personalized responses and foster active student participation in learning \cite{tai2023impact}. Despite challenges in traditional education, such as limited teacher-student interaction and time constraints, chatbots emerged as promising tools to address individual student needs and promote independent learning \cite{fu2020affordances}. In various major fields, including medicine \cite{shinde2021healthcare}, engineering \cite{senthilkumar2019ai}, business \cite{sheehan2020customer}, and education \cite{fryer2017stimulating}, chatbots demonstrate potential benefits and improved academic performance. Recent studies advocate for their integration to assist students in after-class review sessions, aiming to enhance learning outcomes and student motivation through interactive and personalized educational support \cite{vozenilek2004see}. By leveraging AI capabilities using learning theories of constructing knowledge, educators can reimagine learning experiences, moving beyond traditional paradigms to embrace dynamic approaches that cater to individual learner needs and promote deeper engagement. This paradigm shift underscores the transformative potential of GenAI in shaping the future of education, a theme that resonates throughout our exploration of AI-enabled learning in this manuscript.

\noindent Nevertheless, rigorous empirical research is needed to establish the true efficacy of these AI-powered systems in enhancing student learning outcomes compared to traditional methods. This study aims to answer that very question. In this research, we plan to execute a controlled trial focusing on an LLM-based intelligent tutoring system named \textit{"SAMCares: An Adaptive Learning Hub"} (From this point forward, we will refer it to \textit{SAMCares}). The system is developed by leveraging LLaMa-2 70B \cite{touvron2023llama} (a large language model developed \& released by Meta) as the base model. In addition, The system will utilize a fixed database, meticulously created to serve as a reliable knowledge repository. This database contains only course materials taught at SHSU. A Retriever Augmented Generation (RAG) system is used to access this database in real-time and provide context-aware and adaptive educational support. Our central hypothesis assumes that the integration of \textit{SAMCares} into the educational framework will result in significant improvements in several key areas, including but not limited to greater knowledge gains, higher student satisfaction, and lower cognitive load compared to standard educational materials alone. Finally, this study aims to validate the effectiveness of \textit{SAMCares} and contribute valuable insights into the potential of AI-enhanced learning tools in modern educational settings.

\noindent The remainder of the manuscript is structured as follows. The following section describes the overall study design, recruitment plan, and data collection and storage procedure. Next, we describe the building and usage of \textit{SAMCares}. Finally, we provide potential challenges and limitations and end with concluding remarks.


\section{Methods}
\label{Section:Methods}
The primary goal of this research is to investigate whether GenAI tools (especially text generation tools) can enhance students' learning experience and consequently improve their academic performance. To create such a tool \textit{SAMCares} that can address and adapt to the varied educational needs of students, reflecting our commitment to enhancing the quality and accessibility of learning experiences in higher education. Currently, the developed \textit{SAMCares} tool has the following features built-in:
\begin{enumerate}
    \item An interactive and personalized learning experience where students can ask questions, access relevant resources, and seek clarification.
    \item Quick retrieving process through RAG to keep the knowledge base grounded.
    \item Facilitates personalized learning paths, as required by students' needs.
    \item Options to add additional learning resources /materials by students to cater to additional queries they might have.
\end{enumerate}

\noindent Below, we describe the different aspects of the proposed study protocol

\subsection{Study Design and Study Population}
\label{Subsection: Study_Design_Population}
This study will be a year-long, stratified, randomized, controlled trial. The sample size for the study is determined to be 150 (A detailed power study is provided in a later section). The inclusion criteria for the study are: (1) Currently enrolled as an undergraduate student at SHSU; (2) 18+ years old; (3) Freshman; (4) Should possess a basic level of technological proficiency; Exclusion criteria: (1) Unwilling or unable to provide the informed consent forms; (2) Unwilling to complete the assignments; (3) Unavailable during the entirety of the study; (4) Unwilling to attend follow-up surveys or interviews. 

\noindent This study requires collecting and storing data from human subjects for subsequent analysis. The Institutional Review Board (IRB) determined this study doesn't require IRB oversight. 
This plan includes obtaining informed consent from all participants, ensuring they are fully aware of the purpose of the study, and their right to withdraw at any time without any adverse consequences. The collected data will be stored securely in an encrypted data repository. 

\subsubsection{Sample Size for the Pilot Study}
\label{Subsubsection: Sample_Size}
Before performing the study, it is crucial to determine an appropriate sample size to ensure our findings' statistical validity and reliability. To achieve this, we conducted a comprehensive power analysis. This was done to calculate the minimum number of subjects required to detect a significant effect of the tool usage, considering a conventional $\alpha = 0.05$ and a power level of $80\%$.
The expected effect size was derived from preliminary course analysis data and pertinent literature \cite{becker2011integrative}. We hypothesized that participants interacting with \textit{SAMCares} tool would demonstrate superior outcomes, with an expected mean score of $80$ out of $100$ (in quantitative test scores), compared to a mean score of $78$ in the control group, both with a standard deviation of $4$. The power calculation was performed using the following formula-
\begin{align*}
    k &= \frac{n_2}{n_1}\\
    n_1 &= \frac{(\sigma_1^2 + \sigma_2^2/K)(z_{1-\alpha/2}+z_{1-\beta})^2}{\Delta^2} 
\end{align*}
\noindent where, 
absolute difference between two means, $\Delta = |\mu_2 - \mu_1|$, $\sigma$ is the variance of mean, $n$ is the sample size, $\alpha$ is the probability of \textit{Type I} error, $\beta$ is the probability of \textit{Type II} error, critical $z$ value for a given $\alpha$ or $\beta$, and $k$ is the ratio of sample size for groups. For a $1:1$ ratio recruiting, the power analysis indicated that a minimum sample size of 63 participants per group (total recruitment of 126 participants) was optimal for this study. After careful consideration and adjustments for potential dropout rates and population variability, a total of 150 participants will be recruited for this study. This size balances the need for sufficient statistical power and practical considerations such as time, cost, and resource availability.

\subsubsection{Recruitment}
\label{Subsubsection: Recruitment}
Participant recruitment will be kept limited to the freshman class of the SHSU Engineering Technology Department. 
This will be facilitated through (1) referrals from faculty members within the department who have agreed to recommend their students for the study; (2) a targeted email communication will be sent to department students; (3) Physical flyers will be posted in key locations across campus; and (4) Departments social media platforms will be used to extend our recruitment efforts further. 

\subsubsection{Randomization}
\label{Subsubsection: Randomization}
\begin{figure}[H]
    \centering
    \includegraphics[scale = 0.5]{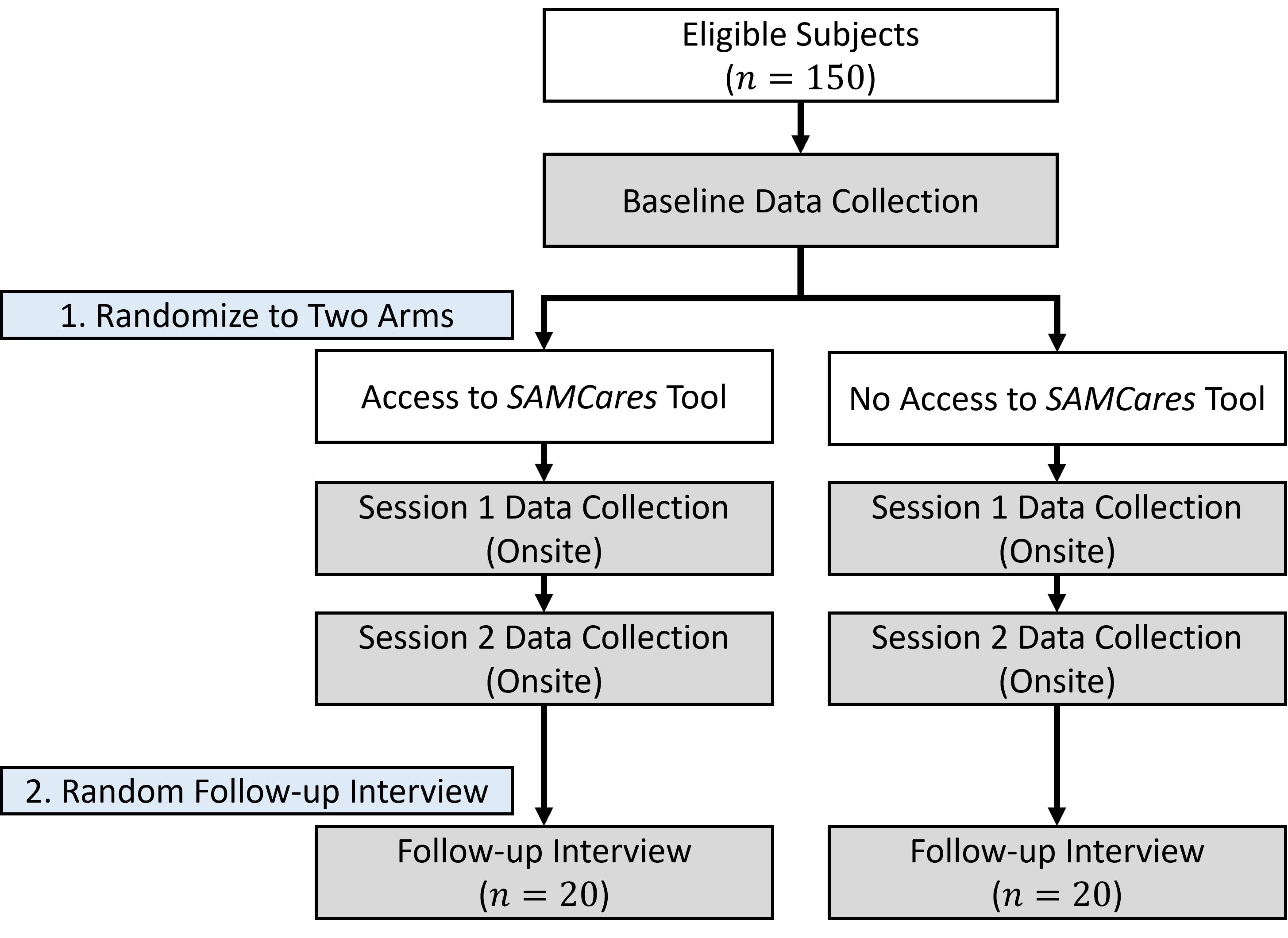}
    \caption{Schematic Representation of Randomization and Data Collection Process for \textit{SAMCares} Tool Evaluation}
    \label{Figure:Randomization}
\end{figure}

Eligible participants will be randomly assigned into two groups; either the group will have access to the \textit{SAMCares} tool ($n = 75$)  or the group without access to the \textit{SAMCares} tool with a $1:1$ allocation. A randomization code will be generated to assign participants to either of the groups. Only after the final selections have been made and the consent forms are signed will the participants be able to know which group they are a part of. A schematic chart of the randomization is shown in Figure \ref{Figure:Randomization}.

\subsection{Data Collection}
\label{Subsection:Data_Collection}
A structured data collection process has been meticulously planned. A wide array of data will be gathered to ensure a comprehensive understanding of the \textit{SAMCares} tool's effectiveness (in general, the effectiveness of GenAI in an educational environment). This includes quantitative data such as exam scores and topic assessment tests, which will objectively measure participants' performance and knowledge retention. 
\noindent Additionally, qualitative data will be collected through pre- and post- study surveys, offering insights into participants' perceptions, satisfaction levels, and subjective experiences with the \textit{SAMCares} tool. An eye tracker will be installed to measure participants' level of engagement and focus during the sessions. This should also reveal the patterns of gaze direction, engagement with the tool, points of focus, and visual fatigue analysis. This will provide us insight into the usability and cognitive load of the \textit{SAMCares} interface.

\noindent The data collection process will also include video recordings. These recordings will add context by documenting participants' behaviors and interactions throughout the study sessions. This will provide an insight into the learning environment, offering a detailed and nuanced perspective on how participants engage with the material and the \textit{SAMCares} tool. A sub-sample of participants will be interviewed for follow-up on the study ($n=20$ from each group). These interviews will also be recorded for further qualitative analysis. The goal of this follow-up interview is to get additional information and experience from the participants.  Some potential questions that will be asked during this interview will include but are not limited to:
\begin{enumerate}
    \item How do the participants describe their experience using GenAI/\textit{SAMCare} compared to traditional learning methods?
    \item In what ways did GenAI/\textit{SAMCare} help them to better understand and engage with the topics, and how did it fall short?
    \item What specific features or functionalities of GenAI/\textit{SAMCare} were most useful or frustrating for them and why?
    \item How did participants emotional and motivational states (e.g., confidence, anxiety, interest etc.) shifted during the use of GenAI/\textit{SAMCare}, and what factors contributed to these changes?
\end{enumerate}

\subsubsection{Trial Sessions}
\label{Subsubsection:Study_Sessions}
Each participant will have to attend a total of two sessions. Each session will be approximately 1 hour long. They will be provided with study materials (see the \textit{Materials and Assessment} section for details). One of the groups will have access to the \textit{SAMCares} tool, while the other group will be limited to the study materials. At the beginning of the study, the participants will complete the pre-survey and an instructional video on how to use the tool will be shown to them before starting the session (only for the participants in the \textit{SAMCares} group). 
The pre-survey forms will collect demographic information (age, gender, race, and other relevant variables, etc.) and pre-session information related to the study. 
The participants will also be shown a set of instructional videos on how to use \textit{SAMCares} the tool effectively. This will include a demonstration of how to effectively ask the same question in different ways to get the best possible explanation that suits their needs (details in \textit{Importance of Question} section).  
The participants will start the session by watching a 15-minute video on a specific topic (randomly assigned for each session) selected beforehand. Afterward, they will be provided with study materials and the \textit{SAMCares} tool. The participants will be given the next 30 - 45 minutes to study the materials provided with or without the help of the \textit{SAMCares} tool. At the end of the study session, the participants will be asked to complete a short assessment test based on the study topic to test their retention and understanding of the topic. A second session will be conducted on a separate day following the same procedure as the first session. Once participants have completed two sessions, they will be asked to complete a post-trial survey. 

\subsubsection{Materials and Assessment}
\label{Subsubsection: Assessment}
For this study, several topics are meticulously selected and designed to fit into a concise, 15-minute teaching segment. The aim is not to cover these topics exhaustively but to give participants an introductory idea of the topic and see how they perform on them. For each selected topic, an array of educational materials will be developed, comprising a short video lecture limited to 15 minutes focused on the main concepts and key points. Alongside the video, participants will have access to lecture notes and the presentation file for review and study purposes. Additional reading materials will be provided to deepen their understanding of the topic. A varied question bank designed for topic assessment will further reinforce the learning process, including multiple-choice, true/false, fill-in-the-gaps, and multiple-answer questions, along with descriptive questions where necessary.

\noindent A structured short assessment test will be implemented for quantitative assessment. These tests will be prepared based on the study topics, and the participants will be introduced earlier. The question formats will follow the same format as the question banks provided to the participants during the study sessions. These assessments will help evaluate the effectiveness of \textit{SAMCares} in enhancing academic achievement. This will be used as the primary outcome of this study. 

\subsubsection{Primary and Second Outcome}
\label{Subsubsection: Primary Outcome}
The study's primary outcome is to assess the effectiveness of GenAI tools like \textit{SAMCares}, which can help improve educational performance. This assessment will be conducted by analyzing the variance in assessment scores, survey analysis, and the magnitude of learning gains between participants granted access to GenAI tools (i.e., \textit{SAMCares}) and those without. Through this multifaceted approach, the study aims to provide a clear and evidence-based understanding of how SAMCares and similar GenAI tools can contribute to advancing the quality of education in modern academic environments.

\noindent The secondary outcomes of the study will be the qualitative analysis of the participants' experiences with \textit{SAMCares}, aiming to gain deeper insights into user satisfaction and areas for improvement. This analysis will primarily utilize data from post-study surveys, focusing on gathering feedback about the tool's design, usability, and overall effectiveness. Additionally, eye-tracking data will help evaluate participants' engagement and fatigue during study sessions. By analyzing patterns in gaze and focus, we can assess participants' cognitive load and attention span, which will inform modifications to make course materials more engaging and less taxing, thereby optimizing the learning experience for future cohorts. 


\subsubsection{Data Repository}
\label{Subsubsection: Data_Repository}
A secure data repository will be set up to store all the collected data during this trial. Only authorized personnel will have access to the repository to ensure the confidentiality and integrity of the data. This repository will be designed to comply with data protection regulations at SHSU and the State of Texas. A de-identified version of the data will also be created and stored in a second repository to share with the scientific community. A request form needs to be submitted to access the data. 

\subsubsection{Documentation and Dissemination Policy}
\label{subsubsection: Documentation_Dissemination_Policy}
To ensure the transparency and reproducibility of this study, each phase of the study will be documented rigorously. Detailed logs of the study sessions will be kept, including dates, times, and any deviations from the standard protocol, along with comprehensive notes on the use of the \textit{SAMCares} tool and the engagement of participants with the study materials.
The trial results will be submitted to peer-reviewed scientific journals, conferences, symposia, and seminars for publication. During these dissemination activities, only aggregated data will be presented/published to ensure confidentiality, with a strict policy against disclosing any identifiable participant data.

\subsection{Building \textit{SAMCares}}
The proposed \textit{SAMCares} utilizes Llama 2 (70B model) \cite{touvron2023llama}, developed by Meta, as its backend model. Llama 2 can be accessed through the official Meta website\footnote{https://ai.meta.com/llama/}, and access to its weights and tokenizers is granted after completing the proper access request forms. This is done to ensure the responsible use of the model. Loading LLaMa 2 on a GPU requires requires 140 GB of memory (70 billion * 2 bytes). Thus, to load it in two smaller GPUs (2 NVIDIA RTX A5000; 8192 Cores; 24 GB Memory) in \textit{SAMCares}, it was quantized to 4-bit precision. Thus, we will need 35 GB of memory (70 billion * 0.5 bytes). Anyone interested in using and creating other applications using LLaMa 2 is suggested to read the provided llama recipes by Meta. This work utilizes some of the suggestions provided by Meta to build \textit{SAMCares}. At the end of the study, a version of \textit{SAMCares} will be open-sourced in GitHub for the research community. 

\begin{figure}[H]
    \centering
    \includegraphics[scale = 0.30]{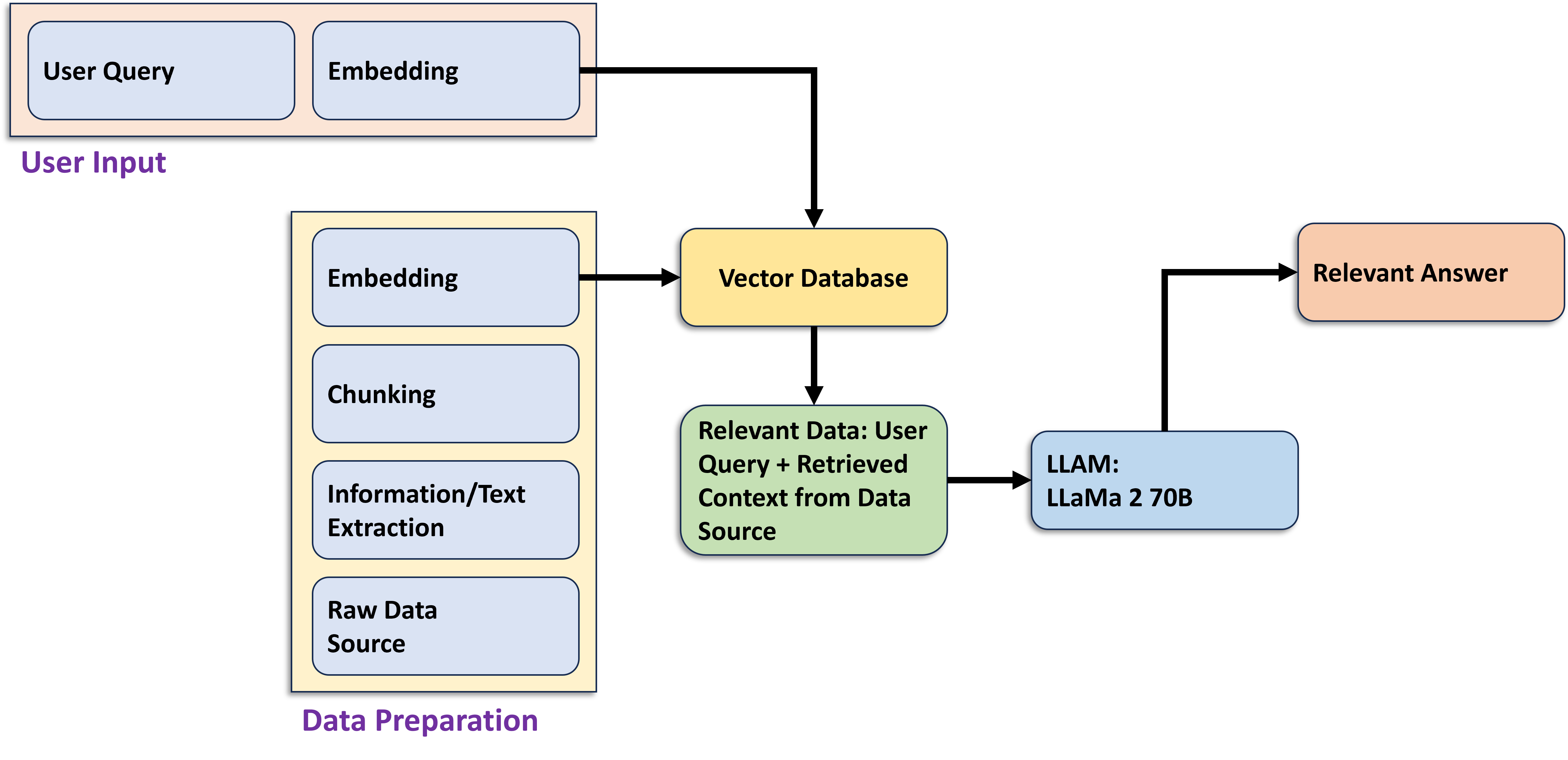}
    \caption{Retriever-Augmented Generation (RAG) system for \textit{SAMCares}: Process flowchart for generating context-aware responses using vector embeddings and LLAMA 2 70B model.}
    \label{Figure: Overall_RAG} 
\end{figure}

\subsubsection{Implementing Retrieval-Augmented Generation in \textit{SAMCares}}
\label{Subsubsection:RAG_SAMCares}
\textit{SAMCares} incorporates a base implementation of the RAG system to augment the LLM's capabilities and minimize hallucinations.  
The RAG approach involves integrating contextually relevant external data sources with the LLM. This data can range from course-specific materials to a wider range of educational resources. RAG requires converting query documents into embedding before passing them into the LLaMa 2 model for querying. Figure \ref{Figure: Overall_RAG} shows the overall process of how \textit{SAMCares} RAG is developed and participants' queries are executed. For more details on RAG, the readers are suggested to read the work by Gao et al. \cite{gao2023retrieval}.

\subsubsection{Web-Based Interface for Demonstration}
\label{Subsubsection: Web_Interface}
\begin{figure}[H]
    \centering
    \includegraphics[scale = 0.6]{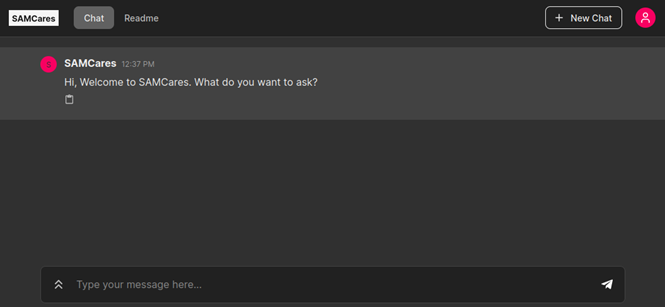}
    \caption{Interface of \textit{SAMCares: An Adaptive Learning Hub} that the participants will be using.}
    \label{Figure: Web_Interface}
\end{figure}

To make \textit{SAMCares} accessible and user-friendly, a web-based interface has been developed. This interface allows participants to interact with \textit{SAMCares} using their SHSU credentials, ensuring secure and personalized access. The interface is intuitive, with features such as a query input field, a response display area, and options to upload additional study materials. Figure \ref{Figure: Web_Interface} shows the interface developed for the usage of this study.

\subsubsection{Importance of Question}
\label{Subsubsection: prompt}
\begin{figure}[htbp!]
    \centering
    \includegraphics[scale = 0.40]{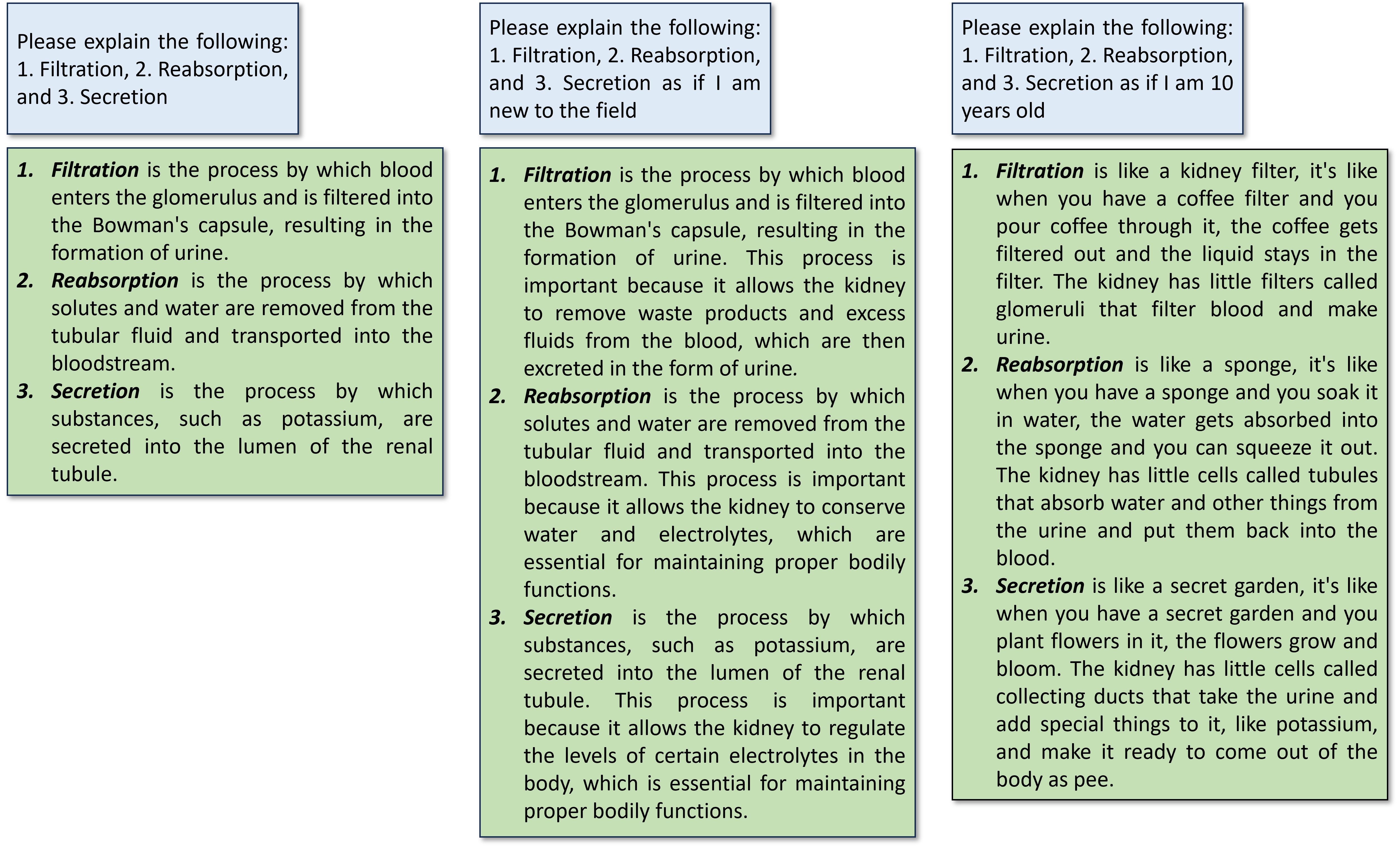}
    \caption{Different levels of explanation from \textit{SAMCares} for the same knowledge, the explanations can change depending on the question asked to the tool.}
    \label{Figure:Propmts}
\end{figure}

In \textit{SAMCares} (or any GenAI model), the role of questions or prompts is pivotal. Effective prompting ensures that the queries directed to the LLM are clear, specific, and helpful in generating useful responses. This is also where adaptive learning comes in. \textit{SAMCares} adapts to users' queries as asked in any manner that can support their understanding. One can ask complex, technical questions or request simplified explanations to make the understanding fit one's needs. This is one of the main reasons to adopt GenAI tools like \textit{SAMCares} as an addition to traditional teaching methods.  Figure \ref{Figure:Propmts} shows such an example where we asked \textit{SAMCares} to explain the processes of filtration, reabsorption, and secretion after uploading a document on \textit{kidney and renal physiology}. Depending on how the question is asked, the first explanation is basic; the second provides more context regarding the importance of each process, and the third offers a metaphorical explanation to make the concepts accessible to a 10-year-old.

\noindent For instance, a query such as, "How does the kidney filter blood? Please explain in detail!" would prompt \textit{SAMCares} to provide a detailed yet understandable explanation tailored to the participant's level of understanding. If a participant requires a deeper understanding, they might ask, "What is the role of reabsorption in maintaining electrolyte balance? Explain in detail." In response, \textit{SAMCares} would offer a more complex explanation involving the relevant physiological processes and implications.
The strength of such a system lies in its capacity to dynamically adjust or adapt to the complexity of its content based on the form of the prompt/question asked. Thus supporting a wide range of educational needs and learning stages. This feature enhances user engagement and promotes comprehension and retention of information, which is essential for an effective learning experience. A guide on how to use prompting and ask questions to get the best possible answers will be provided to this tools users for effective use of the tool. We made \textit{SAMCares} accessible to public for future research and usage. It can be accessed using the following link: \href{https://github.com/SHAFNehal/SAMCares/tree/main}{\textit{SamCares}: An Adaptive Learning Hub}.

\section{Potential Limitations and Challenges}
\label{Section: Limitations_Challanges}
While designing this study we realized there are several potential limitations and challenges that needs to be addressed. One limitation is the potential infrastructure requirements for setting up and running tools like \textit{SAMCares}, which may necessitate access to powerful compute servers or expensive API access through cloud computing platforms such as Amazon AWS or 
Google Cloud. This could create barriers to entry level adoption in resource-constrained academic setup. Another potential challenge will be the risk of misuse by users, who may utilize this tool to complete their homeworks/assignments. This could undermine the educational value of the tool and necessitate the development of safeguards to prevent such misuse. We are currently working on adding fail safes to the current deployed model to detect if it's been asked to complete a homework/assignment and restrict the output. 

\noindent The biggest potential limitation of this study is that it will be a single trial. To validate the results we plan to assess in the future and the widespread adoption of the tool, it is essential to conduct repeated trials in different study sites to establish the tool's effectiveness and generalizability across various contexts. It is also crucial to investigate whether the results are affected by geographical locations, student types (minority students, underserved populations, etc.), or other factors that may impact the tool's performance. This is more so as the current study was conducted on engineering technology students, and it is essential to expand this study to other major departments to ensure the tool's broader applicability and relevance.
\noindent Finally, implementing the tool for students with disabilities poses a significant challenge (from UI/UX design to tool element access). The tool's design and functionality must be adapted to accommodate their special needs, ensuring all students benefit from its capabilities. Addressing these limitations and challenges will be essential to realizing the full potential of this educational tool.

\section{Discussion}
\label{Section: Discussion}
In this manuscript, we documented the protocol to assess the effectiveness and efficacy of GenAI tools in improving student's performance and learning outcomes. The proposed randomized controlled trial aims to provide important empirical evidence on whether emerging GenAI technologies can enhance the learning process when implemented thoughtfully. To the best of our knowledge, this is one of the early efforts to rigorously evaluate GenAI tools like \textit{SAMCares} in an educational context. Our central hypothesis is that integrating tools like \textit{SAMCares} with traditional study materials will result in improved learning outcomes and student satisfaction compared to traditional methods only\footnote{This protocol paper hasn't included any result analysis. We plan to publish the raw data and analyzed results in future publications.}. If confirmed, it would signify the potential of these tools to transform the traditional teaching practices that are designed around student needs. However, careful guidance of AI implementation in education remains critical. The constraints of current GenAI models need careful assessment before wide deployment. There are outstanding questions surrounding optimizing these tools to avoid harm, embed ethical principles, and promote equitable access that require further research.
\noindent Beyond the primary outcomes, our trial will also attempt to uncover insights into how students interact with GenAI tools compared to standard materials. Analyzing usage patterns, queries, focus from eye tracking information, and qualitative feedback will reveal opportunities to optimize these systems for even greater effectiveness. Our findings can inform the iterative development of GenAI tools innovations that drive more equitable, empowering, and learner-centric educational experiences.

\noindent We believe this technology can positively impact students in underrepresented groups in STEM fields, such as women, minorities, students from low-income backgrounds, and students with disabilities. By providing personalized support, the tool may enable these students to overcome traditional barriers to entry and success in engineering and technology disciplines and beyond. Moreover, the tool's interactive and engaging nature should foster increased motivation and interest in STEM subjects among students who may have previously felt disconnected from these fields due to the complexity of explanation and access to study materials. Making tools like this accessible to all types of students, including students with disabilities, may better connect the students in education.

\section{Conclusion}
\label{Section: Conclusion}
This pilot study aims to evaluate the effectiveness of GenAI tools like \textit{SAMCares} as an assistive tool in education. Through its deployment, we will gain empirical data and experiential knowledge of the effect of such tools in attaining students' learning objectives. 
This study will also help identify the broader implications of GenAI in education. By assessing the system's performance across diverse learning scenarios and student populations, researchers and educators can better understand the strengths and limitations of AI as a teaching tool in a way that will provide a framework for future developments in the field, emphasizing the importance of accuracy, user interaction, and adaptability in educational technologies. As we progress, the continual evolution of advanced tools promises the potential to transform traditional learning paradigms and pave the way for a more informed, interactive, and intuitive educational future.

\vspace{4\baselineskip}\vspace{-\parskip} 
\newpage
\footnotesize 
\bibliographystyle{unsrtnat} 
\bibliography{ASEEpaper}


\end{document}